\documentclass{emulateapj} \usepackage{psfig} \usepackage{apjfonts}
\usepackage{amsmath}
\usepackage{natbib}

\newcommand{\myemail}{simone.dallosso@uni-tuebingen.de}

\shorttitle{Gravitational waves from post-merger neutron stars}
\shortauthors{Dall'Osso et al.}

\begin{document}

\title{Gravitational Waves from Massive Magnetars formed in Binary Neutron Star Mergers}

\author{Simone Dall'Osso\altaffilmark{1}, Bruno Giacomazzo\altaffilmark{2,3}, Rosalba Perna\altaffilmark{4}, Luigi Stella\altaffilmark{5}}
\affil{1. Theoretical Astrophysics, University of T\"{u}bingen, auf der Morgenstelle 10, 72076, Germany}
\affil{2. Physics Department, University of Trento, via Sommarive 14, I-38123 Trento, Italy}
\affil{3. INFN-TIFPA, Trento Institute for Fundamental Physics and Applications, via Sommarive 14, I-38123 Trento, Italy}
\affil{4. Department of Physics and Astronomy, Stony Brook University, Stony Brook, NY, 11794,USA}
\affil{5. INAF-Osservatorio Astronomico di Roma, via di Frascati 33, 00040, Monteporzio Catone, Roma, Italy}

\altaffiltext{1}{contact address: \myemail}


\begin{abstract}

  Binary neutron star (NS) mergers are among the most promising
  sources of gravitational waves (GWs), as well as candidate
  progenitors for short Gamma-Ray Bursts (SGRBs). Depending on the
  total initial mass of the system, and the NS equation of state, the
  post-merger phase can be characterized by a prompt collapse to a
  black hole, or by the formation of a supramassive NS, or even a
  stable NS. In the latter cases of post-merger NS (PMNS) formation,
  magnetic field amplification during the merger will produce a
  magnetar and induce a mass quadrupole moment in the newly formed NS.
  If the timescale for orthogonalization of the magnetic symmetry axis
  with the spin axis is smaller than the spindown time, the NS will
  radiate its spin down energy primarily via GWs. Here we study this
  scenario for the various outcomes of NS formation: we generalize the
  set of equilibrium states for a twisted torus magnetic configuration
  to include solutions that, for the same external dipolar field,
  carry a larger magnetic energy reservoir; we hence compute the
  magnetic ellipticity for such configurations, and the corresponding
  strength of the expected GW signal as a function of the relative
  magnitude of the dipolar and toroidal field components.  The
  relative number of GW detections from PMNSs and from binary NSs is a
  very strong function of the NS equation of state (EOS), being higher
  ($\sim 1\%$) for the stiffest EOSs and negligibly small for the softest
  ones.  For intermediate-stiffness EOSs, such as the $n=4/7$ polytrope recently used by Giacomazzo \& Perna or
  the GM1 used by Lasky et al., the relative fraction is $\sim
  0.3\%$; correspondingly we estimate a GW detection rate
  from stable PMNSs of $\sim 0.1-1$~yr$^{-1}$ with Advanced detectors,
  and of $\sim 100-1000$~yr$^{-1}$ with detectors of third generation
  such as the Einstein Telescope. Measurement of such GW signal would
  provide constraints on the NS equation of state and, in connection
  with a SGRB, on the nature of the binary progenitors giving rise to
  these events.

\end{abstract}


\keywords{}

\section{Introduction}

Binary neutron star (BNS) mergers are among the most powerful sources
of gravitational waves (GWs) that are expected to be detected in the
next few years by ground-based detectors, such as advanced LIGO and
Virgo (Abadie et al 2010). BNSs are also the focus of theoretical
modeling of short gamma-ray bursts (SGRBs) since their merger can lead
to the production of relativistic jets and hence generate powerful
gamma-ray emissions (e.g., see Berger 2013 for a recent review). One
of the main scenarios of BNS mergers predicts the formation of a
spinning black hole (BH) surrounded by an accretion torus soon after
the merger, i.e., in less than one second approximately (see Faber and
Rasio 2012 for a recent review of BNS merger simulations).

It is however known that the total mass of the binary, together with
the NS equation of state (EOS), can lead to different dynamics in the
post-merger phase (e.g, see Baiotti et al 2008; Hotokezaka et al 2011;
Bauswein et al 2013; Andersson et al 2013). Depending on the initial
mass of the system, and going from high-mass to low-mass BNSs, the end
result of the merger could be a prompt collapse to BH (e.g., Rezzolla
et al 2010) or the formation of a post-merger NS (PMNS). The latter
could be an {\it hypermassive} PMNS (i.e., supported by strong
differential rotation) which will collapse in less than one second
(e.g., Baiotti et al 2008), a {\it supramassive} PMNS (i.e., supported
by rapid and uniform rotation) and, if the masses are sufficiently low
(e.g., $\sim 1.22 \, M_{\odot}$ in the case of Giacomazzo \& Perna
2013), even a {\it stable} PMNS that will not collapse to a BH
independently of its rotation.

The discovery of two NSs of $\sim 2 \, M_{\odot}$ (Demorest et
al. 2010; Antoniadis et al. 2013) has opened the possibility that
indeed a supramassive, or a stable NS, may be the end result in a
significant number of BNS mergers. Moreover, recent observations of
X-ray plateaus in SGRBs could support the possibility, that at least
in some SGRBs, a supramassive or a stable NS with a strong magnetic
field was formed after merger (Rowlinson et al 2013; see also Metzger
et al. 2008, Dall'Osso et al. 2011). The recent discovery of fast
radio bursts (Lorimer et al. 2007; Thornton et al. 2013) has also been
interpreted as the final signal of a supramassive rotating NS
that collapses to a black hole due to magnetic braking (Falcke \&
Rezzolla 2014; cf. Ravi \& Lasky 2014).
Recent numerical simulations of BNS mergers followed the formation of a stable PMNS with a large mass ($\lesssim 2.36 $ M$_{\odot}$), a relatively large radius ($\approx 15$ km), a spin close to break up and a large degree of differential rotation (Giacomazzo \& Perna 2013). Due to the combined effect of Kelvin-Helmholtz instabilities and dynamo action, a strong amplification of the internal magnetic field occurs promptly and the PMNS settles into uniform rotation, with millisecond spin and a strongly twisted interior magnetic field. The resulting picture is reminiscent of the so-called ``standard" magnetar formation scenario (Duncan \& Thompson 1992), in which
an ultramagnetized NS is formed in the core-collapse of a massive star 
by tapping a fraction of the energy in differential rotation of a fast-spinning proto-NS.

Such a scenario would have a very important role as, along with
explaining some electromagnetic observations, the millisecond
spinning, ultramagnetized PMNS could provide a long-lasting GW signal,
extremely valuable for studying the EOS of NS matter (e.g., Takami et
al 2014), in analogy with what was proposed for magnetars born in the
core-collapse of massive stars (Duncan \& Thompson 1994; Zhang \&
Meszaros 2001; Cutler 2002; Stella et al. 2005; Bucciantini et
al. 2006; Dall'Osso \& Stella 2007; Bucciantini et al. 2008; Dall'Osso
et al. 2009; Metzger et al. 2011).

In this work we address the possible long-lasting GW signal following the BNS merger due to the spindown of a magnetically deformed PMNS,
based on the picture studied by Cutler (2002), Stella et al. (2005) and Dall'Osso et al. (2009). In $\S$ \ref{sec:GW-scenario} we analyse 
the main steps that characterise this picture, and introduce a physical model to calculate the properties of the strongly magnetised PMNS.
In $\S$~\ref{sec:eos} we address the role of the NS EOS in determining
the possible outcome of a merger, and compare expectations with
available data from known BNSs. In $\S$ \ref{sec:detection} we
calculate the strength of the expected GW signals and, based on the
inferred properties of the BNS population, estimate the rate at which
stable or supramassive PMNSs could be detected, relative to the total
population of BNS mergers, by the forthcoming generation of detectors.

\section{A generic scenario for GW emission}
\label{sec:GW-scenario}
The general scenario for efficient GW emission from the newly formed, millisecond spinning and strongly magnetised PMNS
can be summarised as follows (cf. Cutler 2002):
\begin{itemize}
\item The mechanism for field amplification at the merger implies that the axis of symmetry of the strongly twisted magnetic field will start almost aligned with the spin axis.
\item The NS is distorted into an ellipsoidal shape by the anisotropic
  magnetic stress. Free body precession will be excited by even a
  small misalignement between the magnetic symmetry axis and the spin
  axis. We indicate the tilt angle with $\chi$ from here on.

\item Strictly speaking, the largest deformation of the PMNS shape is caused by its fast rotation at $\sim$ kHz frequency. The rotationally-induced distortion is, however, always aligned with the instantaneous spin axis 
and thus plays no role in the dynamics of free body precession (see Cutler 2002 for a detailed discussion of this point). This is why we only consider the magnetically-induced distortion.

\item The energy of freebody precession is viscously dissipated and the conserved angular momentum is redistributed in the stellar interior. As a result,  the PMNS ends up rotating around an axis that corresponds to its largest moment of inertia, so as to minimize spin energy at constant angular momentum.

\item A toroidal magnetic field produces a prolate ellipsoid, {\it i.e.}, one in which the smallest moment of inertia is the one relative to the axis of symmetry of the magnetic field.

 \item For a prolate ellipsoid, viscous dissipation implies that the magnetic symmetry axis is driven orthogonal to the spin axis. This maximises the time-varying quadrupole moment of the rotating top, hence its GW emission efficiency. 
\end{itemize}

After this sequence of events the PMNS becomes a potential source of GWs. The strength of the emitted signal will be determined by the strength of the GW-induced spindown torque, and by the competition with the additional torque due to magnetic dipole braking (see $\S$ \ref{sec:spindown}).

\subsection {Growth of the internal field: theoretical and observational support}
\label{sec:fieldgrowth}
General relativistic MHD simulations of BNS mergers show that a strong
toroidal field is always produced during the merger (Giacomazzo et al
2011), even starting with a purely poloidal magnetic
field. 
For the specific choice of an initial dipole $\sim 10^{12}$ G,
Giacomazzo \& Perna (2013) showed that hydrodynamical instabilities
during the merger can generate poloidal and toroidal components of at
least $\sim10^{13}$ G; further amplification was seen while following
the remnant's evolution for tens of milliseconds after the merger,
with the energy in the toroidal field becoming larger than the poloidal
energy by at least one order of magnitude. It is expected that the
interior field can grow even stronger in the subsequent
evolution, up to $\gtrsim 10^{16} $ G, as suggested by recent local
simulations at very high resolution
(Zrake and MacFadyen 2013; Giacomazzo et al. 2014).

The timing and X-ray emission properties of the galactic population of magnetars suggest the presence 
of internal fields much stronger than the external dipoles (e.g. Dall'Osso et al. 2012). Internal magnetic fields 
of $\sim 10^{15}$ G are also derived by energy arguments, in particular based on the 2004 
Giant Flare from SGR 1806 (Stella et al. 2005). Recent observations (Rea et al. 2010, 2012, 2013) have revealed 
outbursting behaviour and enhanced quiescent X-ray luminosity in a few NSs with dipolar fields in the
$5\times 10^{12}-5\times 10^{13}$~G range, well below the few~$\times
10^{14}$~G strength believed necessary to cause crustal fractures, trigger magnetic outbursts and enhance the 
NS quiescent X-ray luminosity.  By means of magnetothermal simulations of the NS crust it was shown
(Perna \& Pons 2011; Pons \& Perna 2011; Vigan\`o et al. 2013) that magnetic stresses can
fracture the crust even in NSs with relatively low external dipoles, as
long as the internal toroidal field is very strong ($\gtrsim$ a few
$\times 10^{15}$~G), thus accounting for the outbursts of 'low-$B$'
field NSs.

\subsection{The twisted-torus magnetic configuration}
\label{sec:twisted-torus}
The magnetic field of the PMNS at the end of the merger phase is expected to quickly settle into an equilibrium state, driven by the growth of magnetic instabilities on very short 
timescales.

\begin{figure*}
\centerline{
\includegraphics[width=8.2cm]{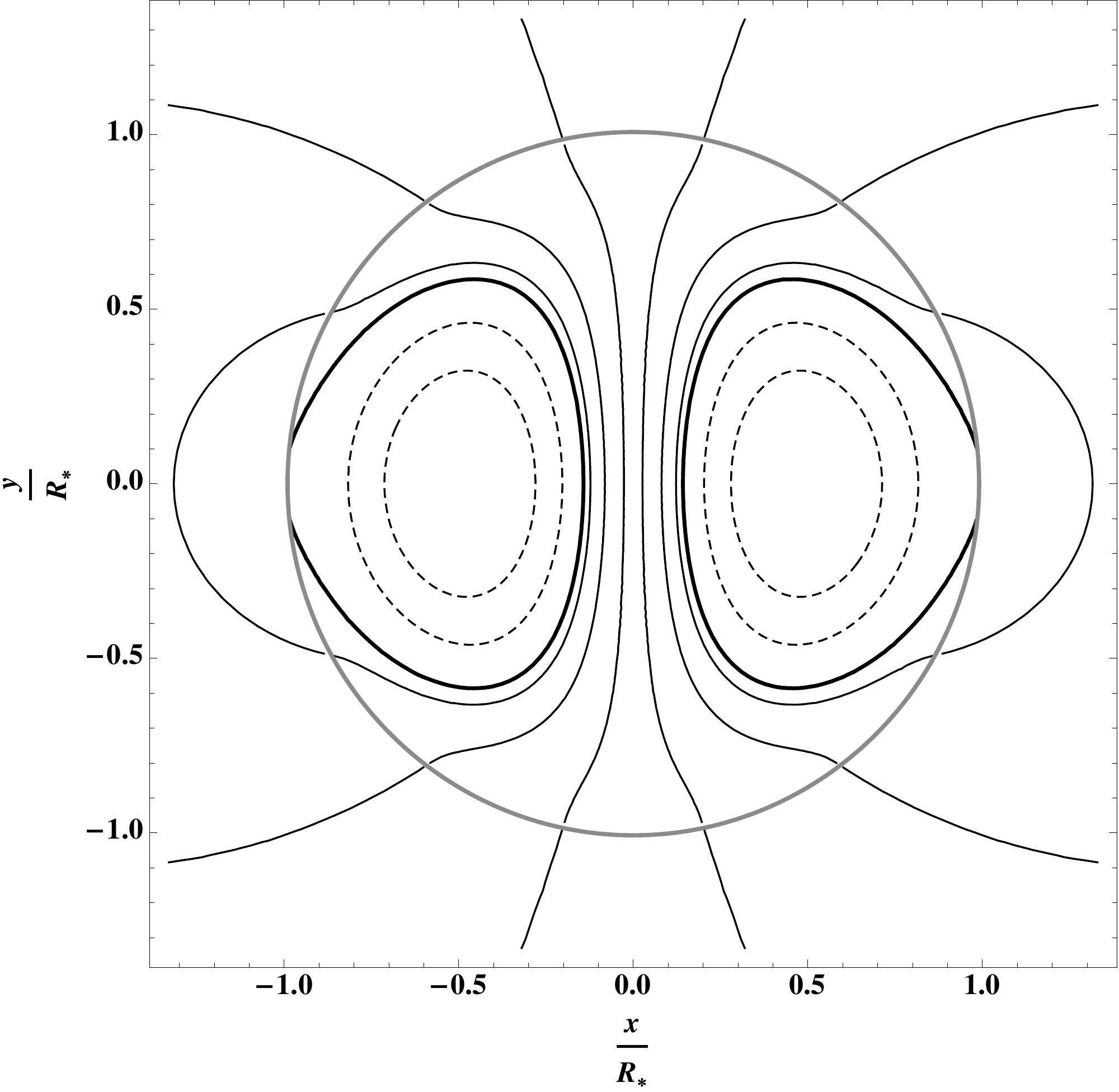}}
\caption{Representation of magnetic field lines in the chosen
  configuration. The NS surface is indicated by the thick gray dashed
  circle. A mixed toroidal-poloidal field in the NS interior is
  matched to a pure dipole in the exterior with no surface
  currents. The toroidal field is confined within the region of closed
  poloidal field lines, the boundary of which is indicated by the
  thick closed curve. The extension of this region can be adjusted
  freely at a fixed strength of the exterior dipole, B$_{\rm
    dip}$. This adjustment induces: {\it i)} a change in the total
  poloidal field energy, without changing B$_{\rm dip}$; {\it ii)} a
  change in the total NS ellipticity, $\epsilon_{\rm B}$, at fixed
  values of B$_{\rm dip}$ and E$_{\rm T}$ (see $\S$
  \ref{sec:ellipticity}); {\it iii)} a change in the stability
  threshold for the toroidal-to-poloidal field ratio (see $\S$
  \ref{sec:stability}).}
\label{fig:field-shape}
\end{figure*}

A general equilibrium configuration for a wide range of initial conditions was found by means of 
extensive numerical simulations (Braithwaite \& Nordlund 2006, Braithwaite 2009) in the form of the
so-called twisted-torus, {\it i.e.} a linked toroidal-poloidal magnetic field.  The poloidal component 
contains an inner bundle of field lines that close inside the NS and therefore do not contribute to 
the exterior field. 

We consider here such a configuration and, following the existing
literature, restrict attention to the case where the toroidal field
does not reach the exterior, which requires that it remains confined
within the close-field-line region.  While only a part of the poloidal
flux extending beyond the NS surface contributes to the exterior
dipolar field, both closed and open poloidal field lines contribute to
the total poloidal energy. The latter will thus depend explicitly on
the size of the closed field line region, and might exceed the energy
of the (exterior) dipole even by a large factor.

In order to derive the relevant physical properties of a magnetised
PMNS we have slightly generalised previous treatments of the
twisted-torus (Mastrano et al. 2011, 2012) to allow for an arbitrary
size of the closed-field-line region, according to the prescription of
Agk\"{u}n et al. (2013). Details of this generalization are provided
in Appendix A1.  We have then chosen a specific configuration of the
magnetic field within a class of solutions that, compared to previous
studies of the twisted torus, allow for {\it i)} a larger magnetic
energy reservoir in the NS interior and {\it ii)} stabilisation of a
stronger toroidal field, {\it for the same strength of the exterior dipole}. 
While our procedure is valid in general, the specific choice of the magnetic 
field geometry determines all numerical estimates. A detailed study of how these
change according to the size of the closed-field-line region will appear in a separate work (in preparation).
%
\subsubsection{Magnetic ellipticity}
\label{sec:ellipticity}
The anisotropic stress due to the interior magnetic field will induce a distortion of the PMNS shape, hence a mass quadrupole moment Q $\sim I_0 \epsilon_{\rm B}$ that is best 
expressed in terms of the moment of inertia of the unperturbed star,
$I_0$, and its {\it total} magnetic ellipticity, $\epsilon_{\rm
  B}$. The latter can be formally defined as the fractional difference
between two main eigenvalues of the moment of inertia tensor. Let the
axis of symmetry of the internal field be the $z$-axis, and the $x$
and $y$-axes lie in a plane perpendicular to it, then $\epsilon_{\rm
  B} \equiv (I_{zz} - I_{xx})/I_0$. To calculate the
magnetically-induced ellipticity we followed Mastrano et al. (2011),
adjusting the calculations to our different choice for the interior
field configuration. More details about our procedure are given in
Appendix A2.

Our main result is the following numerical expression,
\begin{eqnarray}
\label{eq:epsilonb}
\epsilon_{\rm B} &\simeq& 2.725 \times 10^{-6} \left(\frac{{\rm B}_{\rm
      dip}}{10^{14}~{\rm G}}\right)^2 \left(\frac{{\rm R}_*}{15~ {\rm
      km}}\right)^4 \\
&\times&\left(\frac{{\rm M}_*}{2.36~{\rm
      M}_{\odot}}\right)^{-2} \left(1- 0.73~ \frac{{\rm E}_{\rm
      T}}{{\rm E}_{\rm pol}}\right) \, ,
\end{eqnarray}

where the (positive) contribution of the poloidal component and the
(negative) contribution of the toroidal component are consistently
accounted for\footnote{Mastrano et al. (2011, 2012) give $\epsilon_{\rm B}$ vs. 
$\Lambda ={\rm  E}_{\rm pol}/({\rm E}_{\rm pol} + {\rm E}_{\rm T})$. 
The latter goes from 0 (for a purely toroidal field) to 1 (for a purely poloidal field). In terms of $\Lambda$, 
Eq.~(\ref{eq:epsilonb}) becomes $\epsilon_{\rm B} \simeq 4.7 \times 10^{-6}
  \left({\rm B}_{\rm dip} / 10^{14}~{\rm G}\right)^2 \left(1- 0.422 /
    \Lambda\right)$, omitting R$_*$ and M$_*$.}.

\subsubsection{Interior magnetic field vs. exterior dipole}
\label{sec:stability}
Stability considerations set a maximum to the allowed
toroidal-to-poloidal field ratio. Stable stratification of NS matter
allows for much larger values of such ratio than previously thought
(Reisenegger 2009; Akg\"{u}n et al. 2013).  We derived the maximum
allowed ratio for the specific magnetic configuration represented in
Fig.~\ref{fig:field-shape}, by following the procedure of Akg\"{u}n et
al. (2013). Since our discussion here is necessarily limited in scope,
we refer the reader to that paper for a thorough derivation. A
generalisation of our calculations to arbitrary magnetic fields is
postponed to a forthcoming paper.  Adopting the general expression of
the NS field given in $\S$ \ref{Sec:apppend-twistedtorus}, we
integrate its two components within their respective domains (see 
$\S$ \ref{sec:twisted-torus}) and obtain the total energies
\begin{eqnarray}
 \label{eq:def-fields}
  {\rm E}_{\rm pol} & \simeq & 64.675 ~\eta^2_{\rm pol} B^2_0  R^3_{\rm ns} \\
&= & 0.0055 ~b^2_{\rm pol} B^2_0 R^3_{\rm ns} \simeq 5.5 \times 10^{47} \left(\frac{B_{\rm dip}}{10^{14}~{\rm G}}\right)^2 \left(\frac{{\rm R}_{\rm ns}}{15{\rm km}}\right)^3 ~{\rm erg}\nonumber \\
  {\rm E}_{\rm T} & \simeq & 9.152~ \eta^2_{\rm T} B^2_0 R^3_{\rm ns} = 
  0.0105~  b^2_{\rm T} B^2_0 R^3_{\rm ns} \simeq 11.6 \left(\frac{b_{\rm T}}{b_{\rm pol}}\right)^2 {\rm E_{\rm pol}}\, ,
\end{eqnarray}
where $\eta_{\rm pol, T}$ are dimensionless constants measuring the relative strength of the two field components, $B_0$ is the field normalisation, B$_{\rm dip} = 2 \eta_{\rm pol} B_0$ and the following definitions holds: $B^{({\rm max})}_{\rm pol} \equiv b_{\rm pol} B_0$, $B^{({\rm max})}_{\rm T} \equiv b_{\rm T} B_0$, $B^{({\rm max})}$ indicating the maximum value of either field component inside the NS volume.

With these expressions, and after calculating the parameters $k_{\rm
  hydro}$, $k_{\rm pol}$ and $k_{\rm T}$ defined in Eqs.~(79)-(82) of
Akg\"{u}n et al. (2013), we derived the condition for stability of the
magnetic field in terms of the energy ratio between its components
(cf. Eq.~(83) of Akg\"{u}n et al. 2013)
\begin{equation}
\label{eq:ratio-one}
\frac{{\rm E}_{\rm pol}}{{\rm E}_{\rm T}} \gtrsim 0.00894~ \frac{b^2_{\rm T}}{\left(\Gamma / \gamma -1\right) p} \, ,
\end{equation}
where $\Gamma = 1+1/n$ for a polytrope with index $n$, $\gamma$ is the
adiabatic index of the NS fluid\footnote{The factor $(\Gamma/\gamma
  -1) \sim f_p/2 \simeq$ a few \% in a NS core, where $f_p$ is the
  charged particle fraction (Reisenegger \& Goldreich
  1992).}, 
and $p = 8\pi P_c/B^2_0$, with $P_c$ the NS central pressure. The
ratio $b^2_{\rm T}/p$ is derived by inverting
\begin{equation}
\label{eq:ratio-two}
\frac{{\rm E}_{\rm T}}{{\rm E}_{\rm G}} \simeq 0.1098 ~\frac{b^2_{\rm T}}{p} \, ,
\end{equation}
where E$_{\rm G} = \frac{3} {5-n} \frac{{\rm G M}^2}{{\rm R}}$ is the
NS binding energy. Combining Eqs.~(\ref{eq:ratio-one}) and
(\ref{eq:ratio-two}) we finally get
\begin{equation}
\label{eq:stable}
\frac{{\rm E}_{\rm pol}}{{\rm E}_{\rm T}} \gtrsim \frac{0.0814}{(\Gamma/\gamma-1)} \frac{{\rm E}_{\rm tor}}{{\rm E}_{\rm G}} \, .
\end{equation}
Assuming $f_p \simeq 0.05$ (Reisenegger \& Goldreich 1992) the coefficient $\Gamma/\gamma-1 \simeq 0.02$, hence the stability condition reads
\begin{equation}
\label{eq:stabilitycondition}
\left(\frac{{\rm E}_{\rm T}}{10^{50}~{\rm erg}}\right) \lesssim 3 \left(\frac{{\rm B}_{\rm dip}}{10^{14}~{\rm G}}\right)   \left(\frac{\Gamma/\gamma-1}{0.02~}\right)  \left(\frac{{\rm R}_*}{15~{\rm km}}\right)  \left(\frac{{\rm M}_*}{2.36~{\rm M}_{\odot}}\right) \, .
\end{equation}


This suggests that, at the end of the amplification process, a massive
magnetar can be formed with a {\it stable} mixed field dominated by
the toroidal component. The latter can in principle tap the energy
$\gtrsim10^{50}$ erg that was originally in differential rotation,
e.g., for the magnetic configuration of
Fig.~\ref{fig:field-shape}. The strength of the exterior dipole will
be determined by the total energy in the poloidal field {\it and} by
the size of the closed-field-line region.

Finally, by adopting the scalings of Eq.~(\ref{eq:def-fields}) we can write 
the stability condition (\ref{eq:stabilitycondition}) as
\begin{equation}
\label{eq:stability-alternative}
\frac{{\rm E}_{\rm T}}{{\rm E}_{\rm pol}} \lesssim 545 \left(\frac{{\rm B}_{\rm dip}}{10^{14}~{\rm G}}\right)^{-1} \left(\frac{{\rm M}}{2.36~{\rm M}_{\odot}}\right)
 \left(\frac{{\rm R_*}}{15~{\rm km}}\right)^{-2} \left(\frac{\Gamma/\gamma-1}{0.02~}\right) \, .
 \end{equation}
%

 \subsection{Spindown of the newly formed NS}
 \label{sec:spindown}  
 A rotating ellipsoid with the symmetry axis tilted with respect to the spin axis
 by an angle $\chi$ has a GW-induced spindown luminosity (Cutler \&
 Jones 2001 and references therein)
\begin{equation}
\label{eq:GW-spindown}
\dot{\rm E}_{\rm GW} = - \frac{2}{5} \frac{G (I \epsilon_{\rm B})^2}{c^5~}\, \omega^6_s \,{\rm sin}^2\chi\, 
(1+15 {\rm sin}^2 \chi) \,
\end{equation}
where $\nu_s$ is the spin frequency and $\omega_s = 2 \pi \nu_s$. 

Once the prolate ellipsoid has become an orthogonal rotator the GW-induced spindown is maximised and the resulting spindown formula becomes
\begin{equation}
\label{eq:GW-spindwon-orth}
\dot{\rm E}_{\rm GW} = - \frac{32}{5} \frac{G (I \epsilon_{\rm B})^2}{c^5}~ \omega^6_s \,,
\end{equation}
which we will use throughout this work.

When the  additional torque due to the dipole magnetic field is included, the total spin down of the PMNS becomes 
 \begin{equation}
 \label{eq:spindown}
 \dot{\omega}_s = - \frac{B^2_{\rm dip} R^6}{6 I c^3} \omega^3_s - \frac{32}{5} \frac{G I \epsilon^2_{\rm B}}{c^5} \omega^5_s \, ,
\end{equation}
where B$_{\rm dip}$ is the dipole field at the NS pole\footnote{The
  magnetic dipole moment is $\mu_{\rm d} =$ B$_{\rm dip} R^3/2$.} and
R the PMNS radius.

\begin{figure*}
\centerline{
\includegraphics[width=8.7cm]{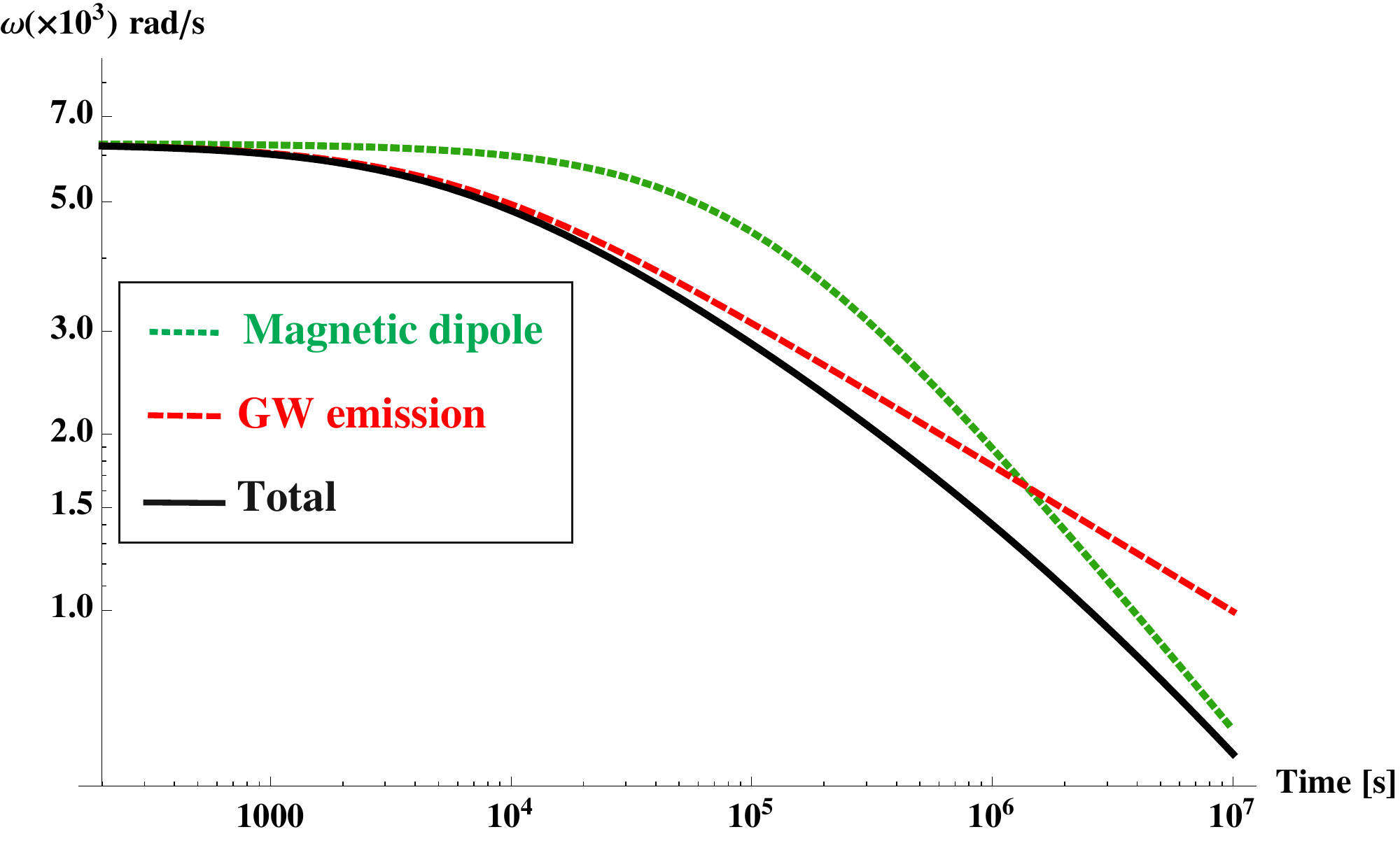}
\includegraphics[width=8.7cm]{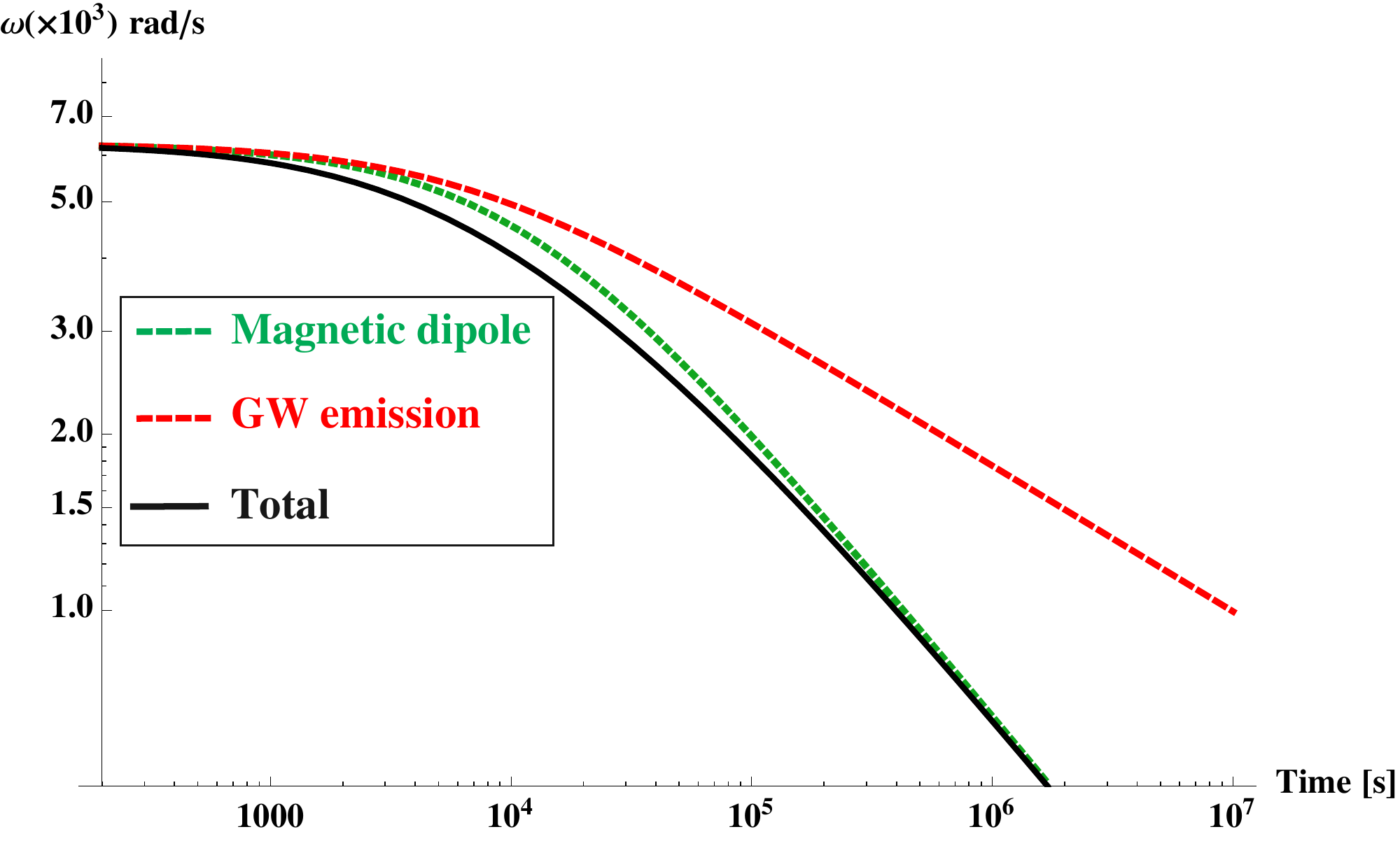}}
\caption{The spindown of a PMNS born with a spin period of 1 ms, in
  two representative cases: {\it Left Panel:} initially the GW-induced
  torque dominates and a strong GW signal can be emitted. The dipole
  magnetic field is B$_{\rm dip} \simeq 10^{14}$ G and the interior
  toroidal field energy corresponds to $\epsilon_{\rm B} \simeq
  10^{-3}$.  As the spin frequency decreases the electromagnetic
  torque progressively kicks in, while the amplitude and frequency of
  the GW signal both decay faster than they would if only GW emission
  were effective. {\it Right Panel:} the spindown is dominated since
  the beginning by the electromagnetic torque. Here B$_{\rm dip}
  \simeq 3\times 10^{14}$ G and the same $\epsilon_{\rm B} \simeq
  10^{-3}$. No strong GW signal is expected in this case, but a bright
  electromagnetic transient, e.g., a short GRB extended emission or
  plateau, could result.}
\label{fig:spindown}
\end{figure*}

In Fig. \ref{fig:spindown} we plot the solution of
Eq.~(\ref{eq:spindown}) in two representative cases, showing the
critical role of the ratio between the interior toroidal field and the
exterior dipole in setting the intensity and duration of the
spindown-induced GW signal.  When GW emission initially dominates the
spindown, it will do so only for a limited time after which magnetic
dipole takes over. If magnetic dipole braking dominates at birth, on
the other hand, it will do so even at later times due to its weaker
dependence on $\omega$. In this case the PMNS spin energy is released
electromagnetically and could produce observables like, e.g. plateaus
in short GRBs (Rowlinson et al. 2013; cf. Dall'Osso et al. 2011 for
long GRBs). 

In this paper we will only be concerned with the GW signal, thus we
aim at determining the conditions under which the GW torque dominates the
PMNS spindown.  However, we will be interested in tracking the
spindown for as long as possible since the population of potential
sources, {\it stable} and supramassive
PMNSs,
will display distinctive features in the evolution of their
signals. The GW signal emitted by a stable PMNS will be characterised
by steadily decreasing frequency and amplitude, with dipole braking
significantly accelerating the evolution at late time. On the other
hand, the collapse of the supramassive object will truncate the signal
thus leaving a very specific signature.

\subsection{Orthogonalization timescale}
Given the near alignment implied by the initial conditions, significant GW
emission will ensue only after the tilt angle $\chi$ has become
large. During this very early phase the PMNS is however subject to
magnetic dipole braking, with a spindown time $\tau_{\rm em,i} =
\omega_i/(2 \dot{\omega}_i) \simeq 1$ day B$^2_{\rm dip,14}$ P$^2_{i,
  \rm ms}$. A necessary condition for the PMNS to be able to radiate
its huge spin energy reservoir via GWs is that the growth time of the
tilt angle, $\tau_{\chi}$, be shorter than $\tau_{\rm em, i}$. In the
opposite case, a bright electromagnetic transient of duration $\sim
\tau_{\rm em,i}$ would carry away much of the initial spin energy,
leaving much less energy available for GW emission once $\chi$ has grown
significantly.

Dall'Osso et al. (2009) derived the expression $\tau_{\chi} \sim
13~{\rm E}_{\rm T,50} {\rm P}^2_{\rm ms} T^{-6}_{10}$ s, where the
strong temperature dependence is due to bulk viscosity being the most
important dissipation mechanism. Using this expression, they calculated
the time for the tilt angle to grow to, e.g. $\pi/3$ rad, explicitly accounting for the fact that the very efficient modified-URCA reactions cause the NS temperature to change significantly during the process.
They concluded that, for the region of parameter space where GW
spindown wins over magnetic dipole braking, orthogonalisation is
always achieved in a time significantly shorter than $\tau_{\rm
  em.i}$.


\section{Stable vs. unstable magnetars: different EOS and time of collapse}
\label{sec:eos}
Whether a merger forms a stable or a supramassive PMNS
will depend on the mass of the binary components and on the maximum
mass (M$_{\rm max}$) allowed by the NS EOS. A stable PMNS can be
formed in the merger of a relatively low-mass BNS (Giacomazzo \& Perna
2013), for a sufficiently stiff EOS that allows a maximum NS mass
${\rm M}_{\rm max} \gtrsim 2.3$ M$_{\odot}$. For a given EOS, fast
rotation\footnote{We only consider uniform rotation here.} provides
additional support against collapse, increasing the mass limit by up
to $\sim$ 20\% when break-up speed is approached (Lyford et
al. 2003). Supramassive PMNSs could thus be formed in a wider range of
conditions and may well represent a large fraction of the whole
population, especially when considering the softer EOS among those
consistent with the observational constraint ${\rm M}_{\rm max} > 2.1$
M$_{\odot}$ .

NS masses generally refer to the gravitational mass, $M_g$, the corresponding rest-mass being approximately\footnote{This accounts for the NS binding energy, including leading-order relativistic corrections as well as finite entropy effects.} (Timmes et al. 1996)
\begin{equation}
\label{eq:merest}
M_r = M_g + 0.075 M^2_g \, . 
\end{equation}
For a given EOS the equilibrium mass of a NS is a function of the central density, 
$\hat{M}_g (\rho_c)$,  and the maximum mass ${\rm M}_{g, {\rm max}}$ indicates the peak in this function. 
Models with $M_g > {\rm M}_{g, {\rm max}}$ are unstable and immediately collapse to BHs when rotation is negligible.

When rotation is included one can formally write the equilibrium mass as a function of the spin period $P$, or of the rotation rate $\Omega$, as (Lasky et al. 2014; Ravi \& Lasky 2014)
\begin{equation}
\label{eq:maxmass-spin}
\hat{M}_g (P; \rho_c) = 
\hat{M}_g (\rho_c) 
+ \Delta M(P; \rho_c) = \hat{M}_g (\rho_c) ( 1+ \alpha P^{-\beta}) \, , 
\end{equation}
where both coefficients $\alpha$ and $\beta$ depend on the star's EOS\footnote{This is true for relativistic models, while in Newtonian models $\beta=2$ and only $\alpha$ depends on the stellar structure.}. The maximum mass,  ${\rm M}_{g, {\rm max}} (P_{\rm min})$, now depends explicitly on the maximum allowed rotation rate, i.e. the mass-shedding limit $\Omega_{\rm max}$ or the corresponding minimum period P$_{\rm min}$. The latter
is given, to an accuracy of a few percent, by (Stergioulas 2003 and references therein) 
\begin{equation}
\label{eq:omegamax}
 \Omega_{\rm max} = {\cal{C}}(\chi_s) \sqrt{G\, {\rm M}_{g,{\rm max}}/ R^3_{\rm max}} \, ,
 \end{equation}
where M$_{g, {\rm max}}$ and $R_{\rm max}$ are the mass and radius of the maximum mass nonrotating model, $\chi_s = 2 G {\rm M}_{g,{\rm max}}/(c^2 R_{\rm max})$ its compactness, and the function ${\cal{C}}(\chi_s) = 0.468 + 0.378 \chi_s$. Comparing with, e.g. the numerical results of Lasky et al. (2014) for three selected EOS, gives indeed a very good agreement.

\begin{figure*}
\centerline{
\includegraphics[width=7cm]{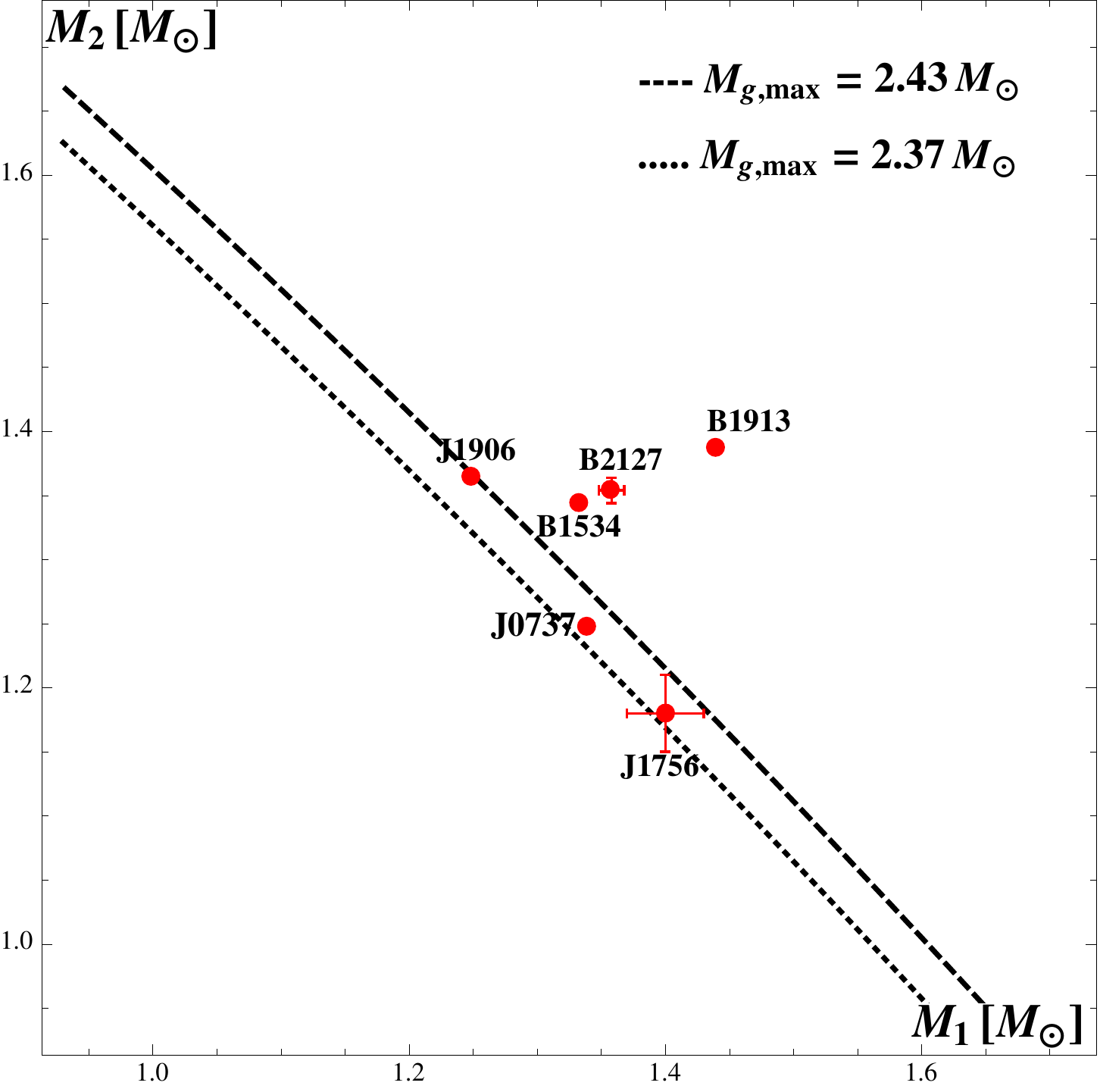}
\includegraphics[width=7cm]{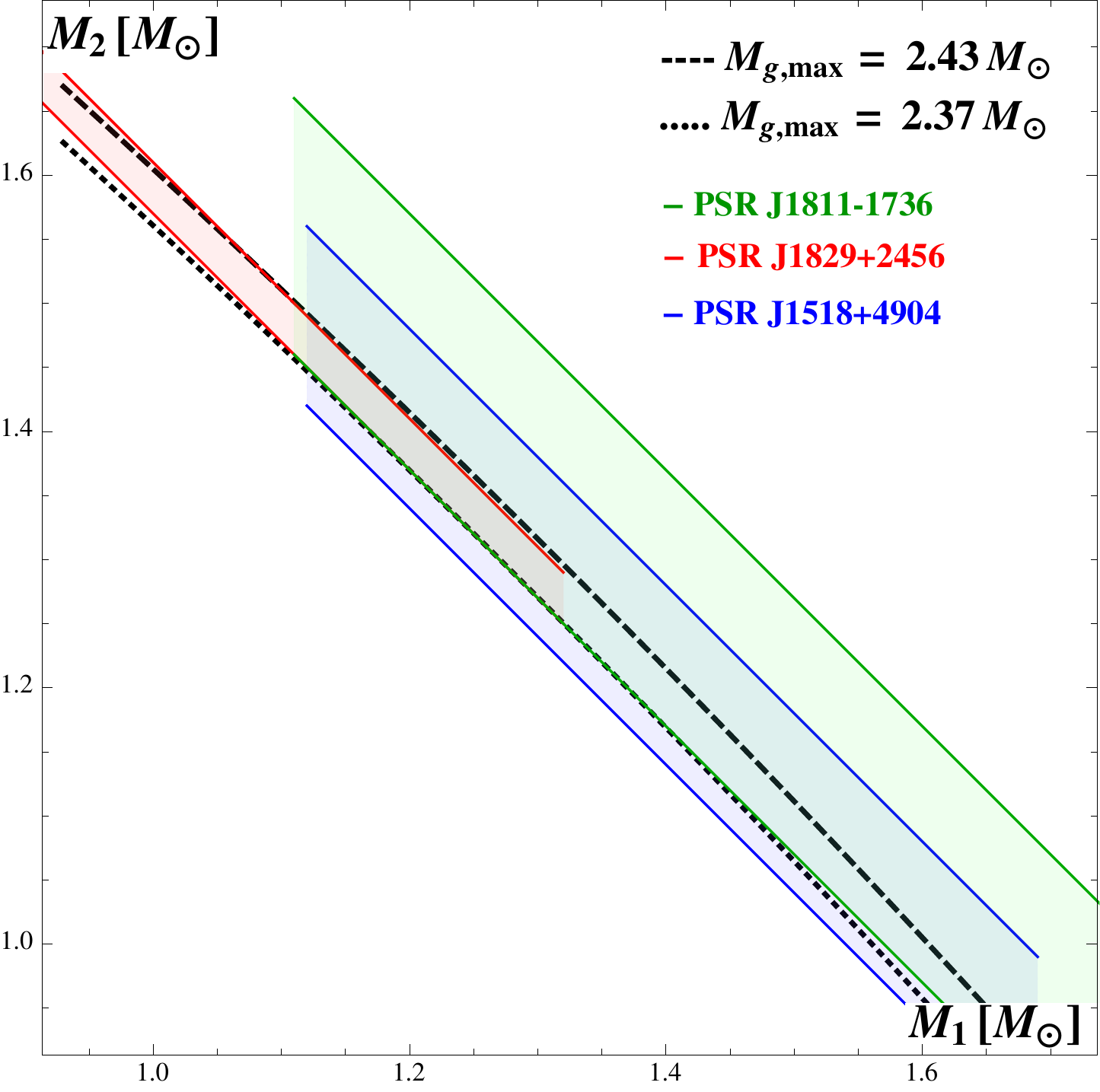}}
\caption{{\it Left Panel:} Measured gravitational masses for NSs in BNSs (Kiziltan et al. 2013). Points below the diagonal lines indicate systems that will potentially form a stable PMNS, {\it i.e.} $M_g < {\rm M}_{g,{\rm max}}$, for two different EOS: {\it a)} a polytrope with $n=4/7$ and $K =30000$, that well approximates the nuclear EOS by Shen et al. 1998 (dashed line); {\it b)} GM1 from Lasky et al. 2014 (dotted line). The two lines are obtained assuming conservation of rest-mass in the merger and an approximate $M_r - M_g$ relation (see text). Mass-loss shifts the lines upwards. For example, the two lines are separated by a rest-mass difference of $\simeq 0.04$ M$_{\odot}$. {\it Right Panel:} For 3 BNSs individual masses are loosely constrained and only the total gravitational mass is well determined. The 1$\sigma$ error ranges (Kiziltan et al. 2013) are plotted: PSR J1829+2456 falls between the two lines. Conclusions are uncertain for the remaining systems.}
\label{fig:masses}
\end{figure*}

The EOS GM1 used by Lasky et al. (2014) has M$_{g,{\rm max}} =2.37 $
M$_{\odot}$ and $R_{\rm max}$ = 12 km, corresponding to $\chi_s \simeq
0.586$ and $\Omega_{\rm max} \simeq 9.3 \times 10^3 $ rad s$^{-1}$
($P_{\rm min} \simeq 0.67$ ms). For this EOS,
$\alpha=1.58\times10^{-10}$ and $\beta=2.84$ (Lasky et al. 2014),
giving ${\rm M}_{g, {\rm max}} (P_{\rm min}) \simeq 2.77$ M$_{\odot}$. For
comparison, Giacomazzo \& Perna (2013) adopted a polytropic EOS with
$n=4/7$ and $K=30000$ that well approximates the behaviour at high
density of the EOS by Shen et al. 1998 (see Oechslin et
al. 2007). This polytropic EOS has a maximum mass M$_{g, {\rm max}} \sim 2.43
M_{\odot}$ (and radius $\sim$ 12 km) for a nonrotating NS, while the
maximally rotating model\footnote{The formulae for rotating models
  give $\Omega_{\rm max} \simeq 9.52 \times 10^3$ rad s$^{-1}$
  (P$_{\rm min} \simeq 0.66$ ms).} has a maximum ${\rm M}_{g, {\rm max}}
(P_{\rm min}) \sim 2.95 M_{\odot}$ (see Giacomazzo \& Perna 2013 for
more details).  Finally, a relatively softer
EOS which is widely used is the APR (Akmal et al.~1998) with ${\rm M}_{g, {\rm max}} = 2.2$
M$_{\odot}$, $R_{\rm max}=10$ km, $\alpha= 3.03\times10^{-11}$ and
$\beta =2.95$ (Lasky et al. 2014). With these figures we get
$P_{\rm min} \simeq 0.51 $ ms and ${\rm M}_{g, {\rm max}} (P_{\rm min})
\simeq 2.54$ M$_{\odot}$.

Based on the measured masses of 9 BNSs, the mass distribution of NSs
in binaries was found to be peaked at $\langle {\rm M}_g \rangle
\simeq (1.32 \pm 0.11)$ M$_{\odot}$ (Kiziltan et al. 2013) which
corresponds to $\langle {\rm M}_r \rangle \simeq (1.45 \pm 0.13)$
M$_{\odot}$, with the errors indicating a 68\% probability interval. A ``typical"
equal-mass binary would have $M_r = (2.91 \pm 0.18) $ M$_{\odot}$, or
$M_g = (2.45 \pm 0.13)$ M$_{\odot}$, close to the maximum for the
$n=4/7$ polytrope described above but uncomfortably large for,
e.g. the APR EOS. Such numbers suggest that, for the $n=4/7$ EOS (or,
possibly, the GM1), a large majority of BNS mergers would produce
either a supramassive or a stable PMNS, with the latter potentially
representing a sizeable fraction. A BH would be the most likely result
for softer EOS, possibly with a small fraction of supramassive PMNSs
rotating close to break-up.

To further clarify this point we plot in Fig.~\ref{fig:masses} the
measured NS masses in 9 BNSs (Kiziltan et al. 2013, their Tab.1) along
with lines indicating ${\rm M}_{g,{\rm max}}$ for the EOS GM1 and the
$n=4/7$ polytrope. These lines assume that the total rest-mass is
conserved in the merger: any loss of rest-mass due to, e.g mass
ejection or the formation of a disk/torus around the remnant, would
shift them upwards in the plot. For three systems only the total
gravitational mass is well determined, hence we plot them separately
showing the 68\% probability range for the total mass in the right
panel of Fig.~\ref{fig:masses}.
\section{Source Detection}
\label{sec:detection}
Three main factors determine the rate at which GW signals of massive
magnetars formed in BNS mergers can be revealed with advanced
detectors: 
{\it i)} the total rate of BNS mergers, $\dot{{\cal{N}}}$, and the fraction of such events
that will form a massive PMNS, call it $p_{\rm ns}$; {\it ii)} the
intrinsic signal strength as a function of the physical properties of
the sources; {\it iii)} the detector's properties.

\subsection{Signal-to-noise ratio}
The maximum strain received from an ideally-oriented\footnote{Ideal orientation to the detector's arms and optimal angle between spin and line of sight.} NS 
spinning at frequency $\nu_s$ and at a distance D is
 \begin{equation}
\label{eq:inst-max-strain}
h(f) = \frac{4 \pi^2 G I \epsilon_{\rm B}}{c^4 {\rm D}} ~f^2 \, ,
\end{equation}
where $f = 2 \nu_s$ is the frequency of the GW signal. 

As the detector collects the signal, the NS spins down and both
frequency and strain decrease. The signal-to-noise ratio for an ideal
matched-filter search
is thus defined as
\begin{equation}
\label{eq:snr}
{\rm S/N } = 2 \left[ \int_{f_i}^{f_f} df \frac{|\tilde{h}(f)|^2}{S_h(f)}\right]^{1/2} \, ,
\end{equation}
where $S_h(f)$ is the detector's (one-sided) noise spectral density 
and $\tilde{h}(f)$ is the Fourier transform of $h[f(t)]$. The latter will depend on the frequency spindown (see Sathyaprakash \& Schutz 2009), hence
on both B$_{\rm dip}$ and $\epsilon_{\rm B}$ in general (cf. Dall'Osso et al. 2009). It is useful for our purposes to write it in the limit where $df/ft$ is only due to GW emission
\begin{eqnarray}
\label{eq:sn-limit}
\left(\frac{{\rm S}}{{\rm N}} \right)_{\rm GW} & \simeq &
10 \left(\frac{{\rm D}}{33.5 {\rm Mpc}}\right)^{-1} \left(\frac{{\rm R}}{15 {\rm km}}\right) \left(\frac{{\rm M}}{2.36 {\rm M}_{\odot}}\right)^{1/2} \nonumber \\
 & & \times \left[ \left(\frac{f_f}{{\rm kHz}}\right)^{-2} - \left(\frac{f_i}{{\rm kHz}}\right)^{-2}\right]^{1/2} \, .
\end{eqnarray}
The choice of the low end of the frequency range, $f_f$, can be very important for the value of S/N, while $f_i$ has a marginal role as long as it is not too close to $f_f$. This will be a crucial point in the next section, where we aim at assessing the effective detectability of our sources.

\subsection{The detector's ``range" ${\cal{R}}$}
The intensity of a received GW signal also depends on the source's direction, and on the orientation of its spin axis with respect to the line-of-sight. At a fixed
detection threshold, favourably oriented sources are detectable out to much larger distances than badly oriented, yet identical, ones. 
A proper   
average of these orientation-dependent horizons, which accounts for the probability of different sources making 
different angles with respect to the detector's arms and having different angles between their spin axis and the line of sight,
is the detector's ``range", ${\cal{R}}$ (Finn \& Chernoff 1993). This allows to write the total rate of detectable events simply
as\footnote{We just added the factor $p_{\rm ns}$ to the formula given by Finn \& Chernoff (1993).}
\begin{equation}
\label{eq:range}
\dot{N}_{\rm det} = \frac{4}{3} \pi \dot{{\cal{N}}} p_{\rm ns} {\cal{R}}^3 \, .
\end{equation}
Note that {\it not all} detected sources will actually be within ${\cal{R}}$: some will be farther away but with a particularly favourable orientation while
others, that are well within ${\cal{R}}$, will go undetected being unfavourably
oriented\footnote{
The fraction of detected sources that will be beyond a given distance can also be estimated
  (Finn \& Chernoff 1993).}.

If we define the optimal horizon D$_{\rm opt}$ as the maximum distance at which an optimally oriented source can be detected with an ideal matched-filter search made using one single interferometer, then the range is simply obtained as ${\cal{R}} = {\rm D}_{\rm opt}/ 2.26$ (see sec. 4.3  in Finn \& Chernoff 1993). 

For a single-detector search we set for simplicity the detection threshold at S/N=8 (Abadie et al. 2010). We don't need to determine it more accurately at this stage, in view of the significant improvement in sensitivity that the operation of a network of detectors will guarantee over the single-detector case (Schutz 2011). 
For this reason, the estimates that follow may well be regarded as conservative ones. 

The maximum distance at which S/N is above threshold, D$_{\rm opt}$,
can only be obtained as a function of source parameters. In
particular, given the dependence of S/N on $f_f$
(Eq.~\ref{eq:sn-limit}), we will consider stable and supramassive NSs
separately. Indeed, while $f_f$ for the former is determined by
spindown causing a decrease of the signal amplitude
(Fig.~\ref{fig:spindown}), for the latter it is determined by the
collapse of the NS which, in general, occurs much earlier (see below).

\subsubsection{Stable PMNS}
\label{sec:stable}
If a stable PMNS is formed in the merger, a strong GW signal starts
within a few tens of minutes, once the angle between the rotation axis
and the symmetry axis of the toroidal field has grown
sufficiently. The initial frequency is $f_i = 2 \nu_{s,i}$ and GWs
should dominate the spindown, initially, in order for the signal to be
detectable at all. As the PMNS spin frequency decreases, the magnetic
dipole torque becomes relatively more important (cfr. Eq.~\ref{eq:spindown}),
the signal amplitude and frequency decay progressively faster than
they would under pure GW emission, and GW emission eventually
fades away once the spin down is dominated by the magnetic
dipole. This determines the lower end of the frequency interval,
$f_f$.

\begin{figure}[t]
\centerline{
\includegraphics[width=8.9cm]{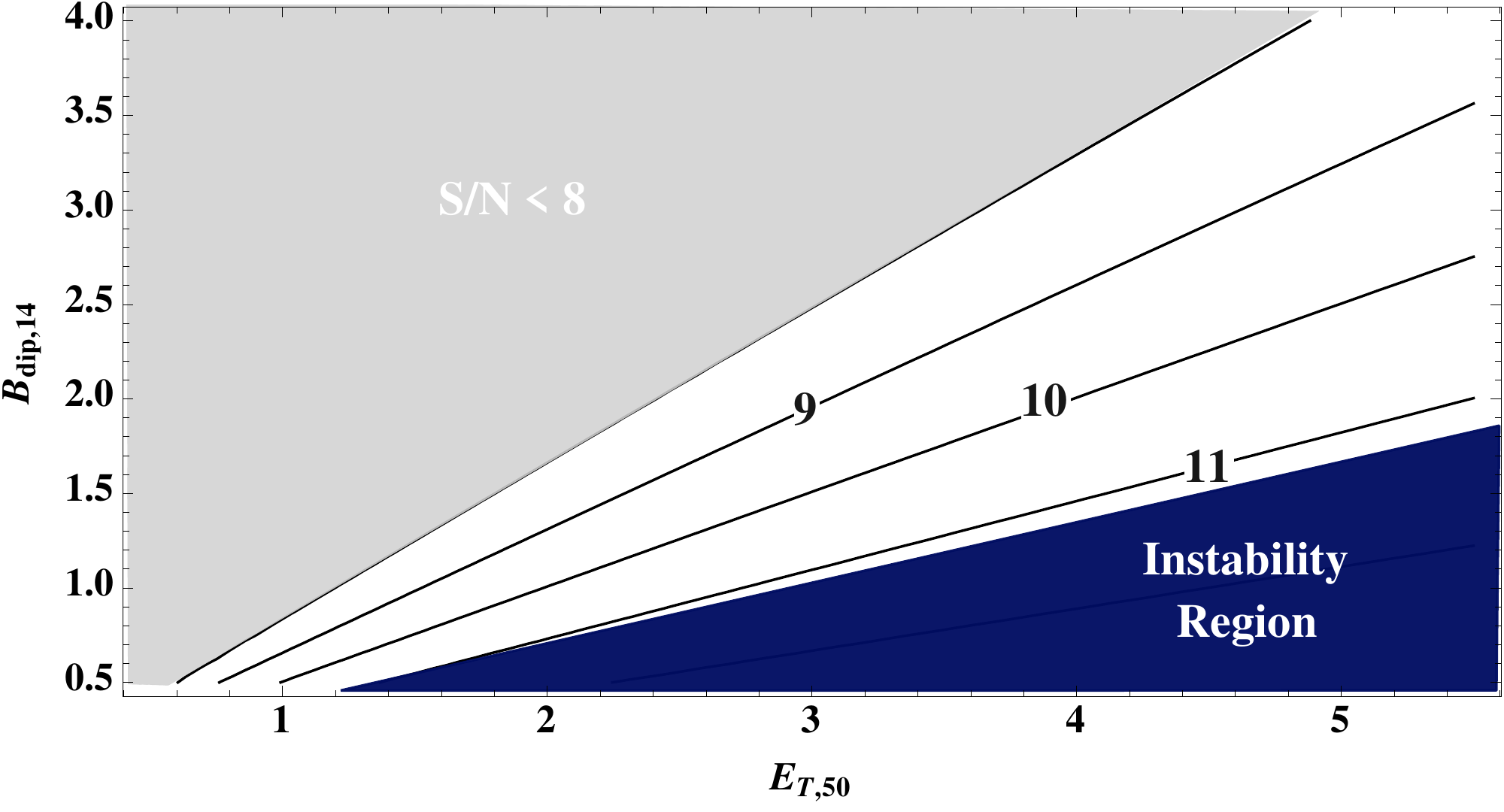}}
\caption{Contours of the optimal signal-to-noise ratio in the B$_{\rm d}$ vs. E$_{\rm T}$ plane, for a single detector search and an ideally oriented source at a distance of 75 Mpc. We define this, somewhat arbitrarily, as the maximum distance at which S/N is above threshold in a sufficiently large region of the parameter space (about half).
Following Abadie et al. (2010), the threshold for detectability is set at S/N = 8. No NS can be found  in the instability region defined by inequality (\ref{eq:stabilitycondition}). }
\vspace{0.1in}
\label{fig2}
\end{figure}

We have calculated S/N according to Eq.~(\ref{eq:snr}) including
self-consistently both torques in the expression for the
spindown. Since S/N depends on two parameters (B$_{\rm dip}$, E$_{\rm
  B}$), the distance up to which it remains above threshold is not
unequivocally determined. We choose as our horizon a distance at which
S/N $\geq 8$ for approximately half of the parameter space of
interest, which turns out to be D$_{\rm opt} \simeq $ 75 Mpc for a
PMNS with M=2.36 M$_{\odot}$, R=15 km and $\nu_{s,i}$= 1 kHz
(cf. Giacomazzo \& Perna 2013), translating to ${\cal{R}} \simeq 33.5$
Mpc. The S/N contours in the B$_{\rm d}$ vs. E$_{\rm B}$ plane for
this specific configuration are shown in Fig.~\ref{fig2}, where the
range of the two magnetic field components was chosen
appropriately for our case.


\subsubsection{Supramassive PMNS}
\label{sec:supramassive}
\begin{figure}[t]
\centerline{
\includegraphics[width=9.5cm]{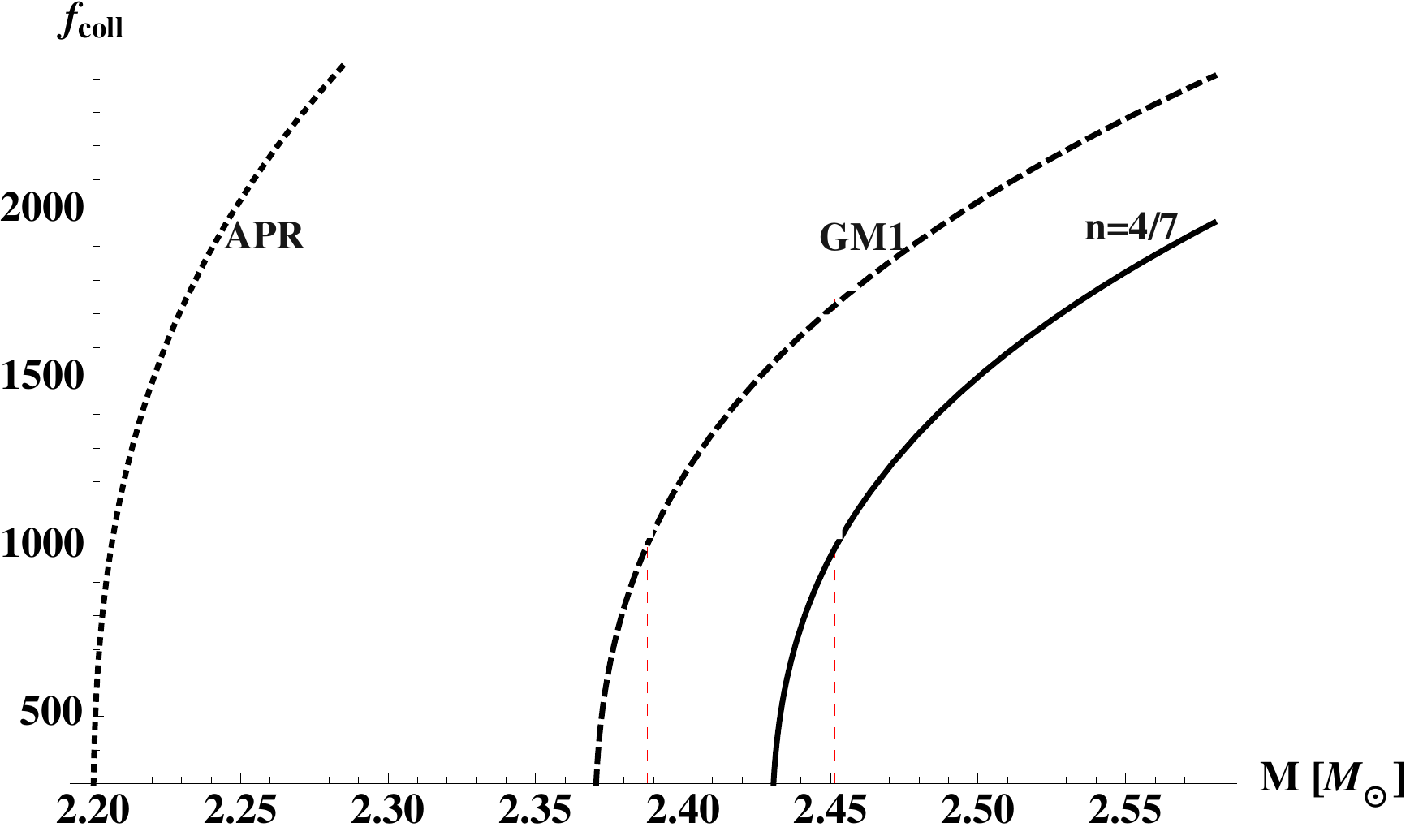}}
\caption{The signal frequency  at which  a supramassive PMNS collapses, $f_{\rm coll} = 2 \nu_{s, {\rm coll}}$ vs. the initial mass, for the three selected EOS discussed in $\S$ \ref{sec:eos}. The red dashed lines indicate the initial mass at which $f_{\rm coll} =$ 1 kHz (spin period of 2 ms), with only lower initial masses collapsing at lower frequencies, implying a narrow range of allowed masses $\Delta$M $\simeq$ 0.02 M$_{\odot}$ for the $n=4/7$ and GM1 EOS. For the APR EOS the allowed range is $\Delta$M $<0.01$ M$_{\odot}$.} 
\label{fig:fcoll}
\end{figure}

The collapse of a supramassive PMNS sets the frequency $ = f_{\rm coll}$. It occurs when the star's mass, M$_{\rm ns}$, equals the maximum mass at a given spin period, ${\rm M}_{g, {\rm max}} (P_{\rm min})$.  By inverting the definition of $\hat{M}_{g} (P; \rho_c)$ of $\S$~\ref{sec:eos} we can write (Lasky et al. 2014)
\begin{equation}
\label{eq:fcoll}
f_{\rm coll} = 2 \left( \frac{{\rm M}_{\rm ns} -  {\rm M}_{g, {\rm max}} }{\alpha {\rm M}_{g,{\rm max}}}\right)^{1/\beta} \, ,
\end{equation}
the solution of which is plotted in Fig.~\ref{fig:fcoll}, as a function of M$_{\rm ns}$, for the three selected EOS discussed in $\S$ \ref{sec:eos}. In general, the frequency at collapse decreases with the mass 
and becomes lower than 1 kHz only if M$_{\rm ns}$ lies in an extremely
narrow range just above ${\rm M}_{g,{\rm max}}$. A much lower S/N than
for stable PMNSs is thus expected, which implies a smaller horizon and
a much smaller number of events. One must restrict attention to the
smallest masses in order to get the strongest signals, visibile to the
largest distances. However, this reduces drastically the number of
possible targets.

\begin{figure}[t]
\centerline{
\includegraphics[width=8.9cm]{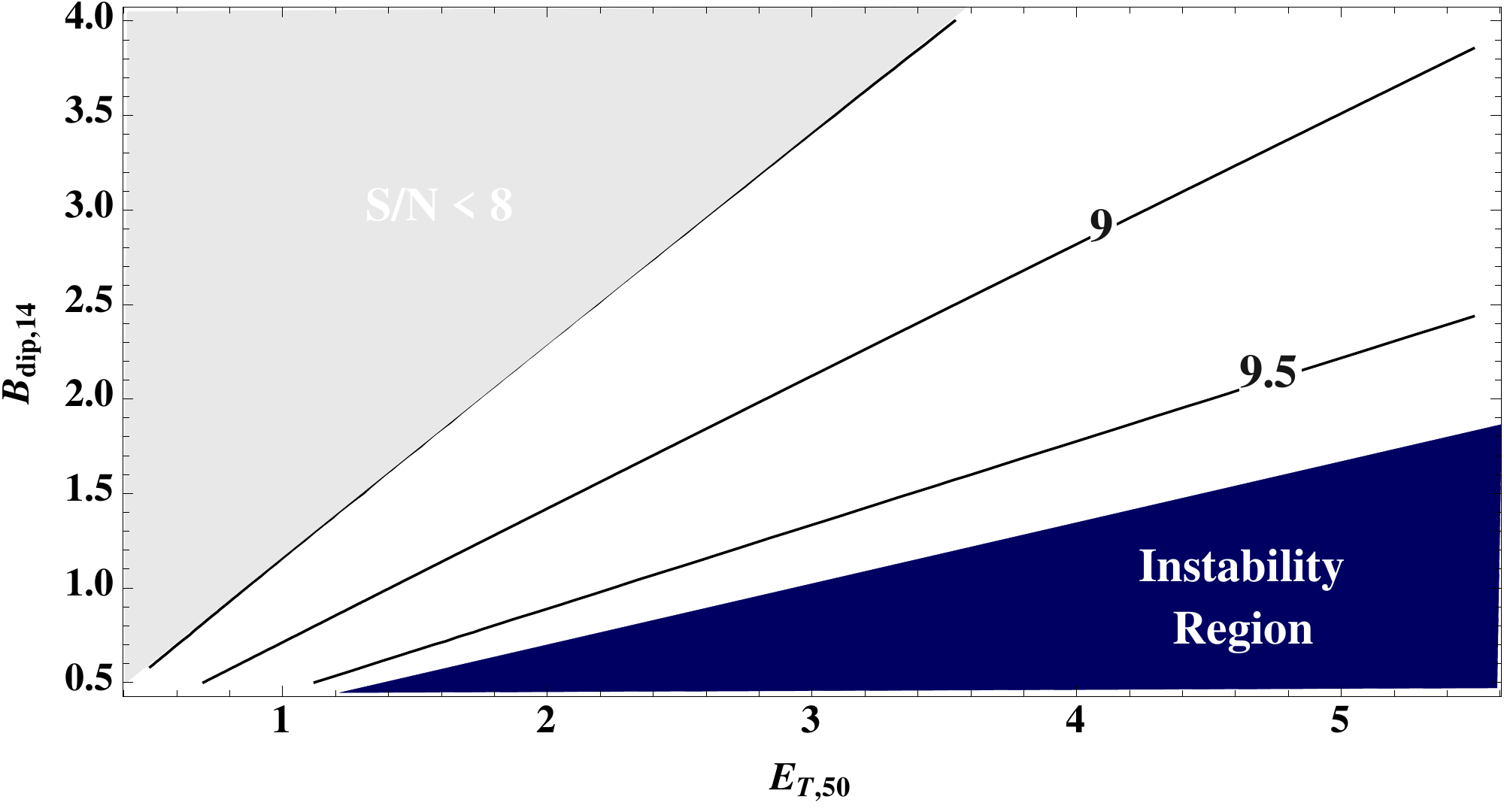}}
\caption{Contours of the optimal signal-to-noise ratio in the B$_{\rm d}$ vs. E$_{\rm T}$ plane, for a single detector search and an ideally oriented {\it supramassive} PMNS at D$_{\rm opt}$ =35~Mpc. We chose M$_g$ = 2.45 M$_{\odot}$, R=15 km and an $n=4/7$ ($K$=3000) polytropic EOS, for which the breakup frequency is 1500 Hz, and the collapse frequency $f_{\rm coll} \simeq 1$ kHz. With the initial spin frequency taken to be at breakup, we have $f_i = 3$ kHz and $ = f_{\rm coll} \simeq $ 1 kHz. We define D$_{\rm opt}$, somewhat arbitrarily, as the maximum distance at which the S/N is above threshold in a sufficiently large region of the relevant parameter space.
  No NS can be found in the instability region defined by 
  (\ref{eq:stabilitycondition}).}
\vspace{0.1in}
\label{fig:Dopt-unst}
\end{figure}

For an approximately Gaussian distribution of remnant masses peaked at 2.45 M$_{\odot}$ and with $\sigma_{\rm M} =0.13$ M$_{\odot}$ (see $\S$~\ref{sec:eos}), only $\sim$ 5 \%  and 6\% of them would lie between ${\rm M}_{g,{\rm max}}$ and (${\rm M}_{g,{\rm max}} + 0.02)$ M$_{\odot}$, for the GM1 and the $n=4/7$ EOS, respectively. This fraction becomes quickly negligible for softer EOS, while for stiffer EOS most mergers would produce stable PMNSs given that ${\rm M}_{g,{\rm max}}$ is above the Gaussian peak. For illustration we have considered a 2.45 M$_{\odot}$ remnant  with R=15 km and initially spinning at break up, $\nu_{s,i} \simeq 1500$ Hz for the $n=4/7$ polytrope. According to Eq.~(\ref{eq:fcoll}) it will collapse when $f_{\rm coll} \simeq $ 1 kHz, or the spin period $\simeq 2$ ms. We derived the optimal horizon for this relatively favourable case as was done in the previous section. The result is shown in Fig.~\ref{fig:Dopt-unst} with D$_{\rm opt} = 35$ Mpc, corresponding to ${\cal{R}} \simeq 16$ Mpc. This considerably smaller horizon compared to the stable PMNSs causes a factor 10 reduction in the sampling volume. Together with the small fraction of objects that fall in this favourable mass range, it implies that a number $\sim$ 100-200 times smaller of such events can be detected compared to the stable PMNSs. 

 

\subsection{The expected event rate}
In light of the above findings, the number of PMNSs that can be
revealed with the forthcoming generation of GW detectors will be
dominated by stable PMNSs, and will strongly depend on the NS EOS.  The
discussion of $\S$ \ref{sec:eos}, summarised in Fig. \ref{fig:masses},
suggests that a sizeable fraction, $p_{\rm ns} \sim 0.2-0.5$, of the
whole population of BNS mergers could result in a stable or marginally
supramassive PMNS for the GM1 or the $n=4/7$ polytropic EOS. This
fraction grows to $\sim$ unity for harder EOS's, in particular those
with M$_{g, {\rm max}} \geq 2.5$ M$_{\odot}$. For relatively softer
EOS's like, e.g. the APR, on the other hand, the maximum mass becomes
quickly too low and essentially all mergers would immediately produce
a BH. The detection of a $\sim$ kHz frequency GW signal with hour-long
spindown following a BNS merger, as discussed here, would thus provide
a very interesting constrain on the EOS of NS matter. This would be
especially valuable when combined with an independent determination of
the total population of BNS, which could provide a direct measure of
the fraction $p_{\rm ns}$.

Since $\dot{{\rm N}}_{\rm det} \propto \dot{{\cal{N}}}$ (cf. Eq. \ref{eq:range}), we can express the rate of detection of PMNS signals relative to the rate of
detection of BNS mergers with Advanced detectors as
\begin{equation}
\label{eq:relative-rate}
\frac{\dot{{\rm N}}_{\rm det}}{\dot{{\cal{N}}}} \sim p_{\rm ns} \left(\frac{{\cal {R}}_{\rm PMNS}}{{\cal{R}}_{\rm BNS}}\right)^3 \sim 0.003 \, ,
\end{equation}
where ${\cal {R}}_{\rm BNS} \sim 170$ Mpc (Abadie et al. 2010) and we assumed $p_{\rm ns} \sim 0.3$ (thus excluding EOSs softer than the GM1). In this case, the small fraction mostly reflects the difference in the sampling volume for the two different types of signals.

``Realistic" estimates of the detection rates of BNS mergers with Advanced detectors range from 40 to 400 events per year (Abadie et al. 2010), and were
derived phenomenologically based on a statistical study of the population of known BNS in the Galaxy
(Kalogera et al. 2004). 
By adopting these numbers, we conclude that stable PMNSs may be detectable at a rate
$\dot{{\rm N}}_{\rm set} \sim (0.1 -1)$ yr$^{-1}$ with Advanced detectors, for the GM1 or the $n=4/7$ polytropic EOS. 
For stiffer EOS, with M$_{g, {\rm max}} \gtrsim 2.5$ M$_{\odot}$, nearly all mergers would produce a PMNS, thus the detection rate could be higher by a factor $\approx 3$.
For softer EOS, on the other hand, these figures drop significantly following the drop in the coefficient $p_{\rm ns}$.

We finally note that third generation detectors, such as the Einstein Telescope, will have  a higher sensitivity by up to a factor $\sim 10$ (Punturo et al 2010). This will increase the detector's range, ${\cal{R}}$ by the same factor, making stable PMNSs detectable up to ${\cal R} \gtrsim 300$ Mpc, and even supramassive ones up to ${\cal R} \gtrsim 150$ Mpc. The detection rate will thus increase by a very large factor, $\sim 10^3$, which is extremely important in particular for supramassive PMNSs. Indeed, with Advanced detectors these
objects will also become interesting sources, with a likely rate of detection of a few events per year.

\section{Summary and Discussion}

The GW signatures of a newly born NS are very sensitive to the
equation of state.  Intense GW emission is expected under the presence
of a strong toroidal magnetic field, as a result of the star's prolate
ellipsoidal shape. In such a configuration, viscous dissipation drives the
magnetic symmetry axis orthogonal to the spin axis, hence maximizing
the strength of the emitted GW radiation.

GWs from highly magnetized NSs newly born in core collapse supernovae
have been studied in a number of works. Here, motivated by recent
numerical simulations of binary NS mergers which show magnetic field
amplification, we have studied the conditions under which strong GW
emission is expected in the post-merger phase, if this is
characterized by the presence of a short-lived, or stable, highly
magnetized NS. To this aim, we have extended the set of equilibrium
states for a twisted torus magnetic configuration to include solutions
that, for a given external dipolar field, carry a larger magnetic
energy reservoir. We have then computed the magnetic ellipticity for 
such configurations, hence the strength of the GW signal,
once the system has orthogonalized.

We find that the strength of the signal, and hence its detectability,
is mainly dependent by the NS EOS. The dependence is twofold.
Firstly, whether the merger of two NSs leads to a supramassive NS
(which eventually collapses to a BH) or to a stable NS, is highly 
dependent on the NS EOS.  For
a given distribution of remnant masses, stiffer EOSs yield a higher
fraction of stable NSs.  Second, the GW signal itself, and in
particular the two distinct and robust spectral features which
characterize the postmerger emission, are very sensitive to the NS EOS
(e.g. Takami et al. 2014).  For an intermediate EOS, such as the
$n=4/7$ polytrope, or the GM1 used by Lasky et al. (2014), we estimate 
that we expect GW emission from PMNSs in about 0.3\% of all 
GW detections from BNS mergers. Correspondingly, we expect a detection rate 
of about 0.1-1 event per year with Advanced detectors; this rate would increase by a factor of $\sim 10^3$ with
third generation detectors, such as the Einstein Telescope, which
will have a higher S/N by up to a factor of 10. These detectors would
be able to observe even the weaker emission from the unstable PMNSs,
albeit with lower rates.

GWs from highly magnetized PMNSs produced in mergers would be
especially interesting if detected in connection with a short GRB.  In
fact, while there is plenty of circumstantial evidence that these
events are produced by a merger, whether the final product is a stable
(or unstable) NS, or a promptly-formed BH is still a subject of
investigation.  Extended emission, occasionally in the form a plateau,
has been seen in about half of {\em Swift} SGRBs.  In some cases, this emission ends abruptly (possibly
indicating the collapse of a hypermassive NS to a BH), while in some
other cases it declines with a powerlaw, possibly indicating the
presence of a stable NS (Rowlinson et
al. 2013). Detection of GWs from the post-merger NS
would allow to discern its identity, and hence shed light on the
nature of the binary progenitors of SGRBs.  In addition, contemporary
detection of a SGRB and GWs would further constrain the NS EOS
(e.g. Giacomazzo et al. 2013).

\acknowledgements For this work S.D. was supported by the
SFB/Transregio 7, funded by the Deutsche Forschungsgemeinschaft (DFG).
B.G. acknowledges support from MIUR FIR Grant No. RBFR13QJYF, and
R.P. from NSF grant No. AST 1009396 and NASA grant No. NNX12AO67G.


\appendix

\subsection{A1: The twisted-torus configuration}
\label{Sec:apppend-twistedtorus}
In spherical coordinates, the interior poloidal field is
\begin{equation}
\label{eq:def-Bint}
 B_{\rm pol} (\hat{r}, \theta) = B_0~ \left[\eta_{\rm pol} \nabla \hat{\alpha}(\hat{r},\theta) \times \nabla \hat{\phi} + \eta_{\rm T} \hat{\beta}(\hat{r}, \theta) \nabla \hat{\phi} \right] \, ,
 \end{equation}
where $\hat{r} = r/R_{\rm ns}$ is the dimensionless radial coordinate and $\nabla \hat{\phi} = \hat{\phi}/(\hat{r}~{\rm sin}\theta)$,  with $\hat{\phi}$ the unit vector in the $\phi$-direction. $B_0$ is a normalisation (in Gauss), $\hat{\alpha}$ the (adimensional) flux function, {\it i.e.} the poloidal magnetic flux threading a polar cap of radius $\tilde{\omega} = R_* \hat{r} ~{\rm sin}\theta$, $\eta_{\rm pol}$ and $\eta_{\rm T}$ are dimensionless constants measuring the relative strength of the two field components and the ``current function" $\hat{\beta} \equiv \hat{\beta}(\hat{\alpha})$, as required by axysimmetry. Since no currents exist in the exterior, this implies that electrical currents can only flow on poloidal field lines that close {\it inside} the NS, hence the bounding region for B$_{\rm T}$. In particular, $\hat{\beta} = (\hat{\alpha} -1)^n$ is usually assumed, with $n>1$ to ensure regularity of the supporting currents at the boundary of the toroidal field region. Finally, the exterior dipole field is matched at the NS surface to the interior field by taking the flux function $\hat{\alpha} (\hat{r}, \theta) = f(\hat{r})~{\rm sin}^2 \theta$. 

Th function $f$ is determined\footnote{A full derivation is given by Akg\"{u}n et al. (2013).} by first imposing the magnetic force and current density to remain finite everywhere inside the NS. For a trial form $f(\hat{r}) \propto \hat{r}^p$ this implies either $p=2$ or $p>3$, suggesting to seek a polynomial solution for $f$. Smoothly matching the interior and exterior fields requires  continuity of the magnetic field at the NS surface, and that no surface currents exist. To satisfy this at least three terms in the polynomial are needed, therefore the ``simplest" solution is $f(\hat{r}) = c_2 \hat{r}^2+c_4 \hat{r}^4 + c_5 \hat{r}^5$. The coefficients are determined by normalisation of $f(\hat{r})$, thus fixing the field shape. Different choices for the polynomial terms are however possible, which affect the shape and size of the closed-field-line region.
For the purpose of this work we have chosen the configuration represented in Fig. \ref{fig:field-shape}, which corresponds to $f(\hat{r}) = (435/8) \hat{r}^2 - (1221/4) \hat{r}^4 + 400 \hat{r}^5 - (1185/8) \hat{r}^6$ and $\hat{\beta} = (\hat{\alpha}-1)^2$. 

\subsection{A2: Magnetic ellipticity}
\label{sec:append-ellipticity}

The two components of the inertia tensor $I_{zz}$ and $I_{xx}$ are obtained through the magnetically-induced density perturbation by (Mastrano et al. 2011)
\begin{equation}
\label{eq:inertia}
I_{jk} = R^5_* \int d{\rm V} \left[\rho(\hat{r}) + \delta \rho(\hat{r}, \theta) \right] (\hat{r}^2 \delta_{jk}- \hat{x}^2_{jk}) \, .
\end{equation}

With these definitions the total ellipticity is eventually expressed as
\begin{equation}
\label{eq:define-epsilon}
\epsilon_{\rm B} = \frac{\pi R^5_*}{I_0} \int d \theta d \hat{r} ~\delta \rho (\hat{r}, \theta)~ \hat{r}^4 ~{\rm sin}~\theta (1- 3 {\rm cos}^2 \theta) \, ,
\end{equation} 
hence the relation between the magnetic field structure and the induced ellipticity of the NS is obtained directly from $\delta \rho (\hat{r}, \theta)$. To calculate the latter we follow the steps described by Mastrano et al. (2011), who write the equation of hydrostatic equilibrium to first order in the magnetic perturbation in the Cowling approximation,
\begin{equation}
\label{eq:hydro-perturbed}
 -\frac{B^2_0}{\hat{r}^2 {\rm sin}^2\theta} \left(\eta^2_{\rm pol} \nabla \hat{\alpha} \hat{\Delta} \hat{\alpha} +\eta^2_{\rm T} \hat{\beta} \nabla \hat{\beta}\right) = \nabla \delta p + \delta \rho \nabla \Phi \, .
\end{equation}
Here $\delta p$ is the magnetically-induced pressure perturbation, $\Phi$ the unperturbed gravitational potential and the Grad-Shafranov operator is $\hat{\Delta} = \partial^2_r + \left({\rm sin}\theta/\hat{r}^2\right) \partial_{\theta} \left[({\rm sin}\theta)^{-1} \partial_{\theta}\right]$.
The $\theta$-component of Eq.~(\ref{eq:hydro-perturbed}) relates the magnetic term to $\delta p$ alone. Feeding the latter in the $\hat{r}$-component gives the density perturbation inside the NS as a function of $\hat{\alpha}$, $\hat{r}$, $\theta$ and the parameters ($B_0, \eta_{\rm T}$, $\eta_{\rm pol}$). The poloidal and toroidal field contribute with opposite signs to $\delta \rho (\hat{r}, \theta)$, hence
the ellipticity due to the poloidal field is positive according to our definition, while the toroidal field produces a negative $\epsilon_{\rm B}$. Solving eq. \ref{eq:hydro-perturbed} for $\delta \rho (\hat{r}, \theta)$ with our chosen  $\hat{\alpha}(\hat{r}, \theta)$, we plug it into eq. \ref{eq:define-epsilon} to eventually obtain the {\it total} magnetic ellipticity of the NS
reported in Eq.~(\ref{eq:epsilonb}).

\end{document}